\documentclass[final,5p,times,twocolumn,authoryear]{elsarticle}
\usepackage{url}
\usepackage{float}
\usepackage{hyperref}
\usepackage{multirow}
\usepackage{booktabs}
\usepackage{enumitem}
\usepackage{threeparttable}
\usepackage{siunitx}
\usepackage{lipsum}
\usepackage{lineno}
\usepackage{amsmath,amssymb,amsfonts}
\usepackage{graphicx}
\usepackage{algorithm}
\usepackage{algorithmic}
\usepackage{array,tabularx}

\begin{document}

\begin{frontmatter}

\title{FacilityAudit-Agent: Beyond Mention-as-Usage for Fine-Grained Facility Contribution Attribution in Astrophysics Publications}

\author[1,2,3]{Wujun Shao}
\author[1,2,3]{Dongwei Fan\corref{cor}}
\author[4,2,6]{Daoqi Peng}
\author[1,2,3]{Yunfei Xu}
\author[1,2,3]{Chenzhou Cui\corref{cor}}
\author[1,2]{Xin Lyu}
\author[5]{Wei Huang}

\cortext[cor]{Corresponding authors\\\textit{E-mail addresses:} \url{fandongwei@nao.cas.cn} (Dongwei Fan), \url{ccz@bao.ac.cn} (Chenzhou Cui)}

\address[1]{National Astronomical Observatories, Chinese Academy of Sciences, Beijing 100101, China}
\address[2]{University of Chinese Academy of Sciences, Beijing 100049, China}
\address[3]{National Astronomical Data Center, Beijing 100101, China}
\address[4]{National Science Library, Chinese Academy of Sciences, Beijing 100190, China}
\address[5]{School of Computer Science, Beijing University of Posts and Telecommunications, Beijing 100876, China}
\address[6]{Groupe d'\'{E}tude des M\'{e}thodes de l'Analyse Sociologique de la Sorbonne, Sorbonne Universit\'{e}, CNRS, Paris 75017, France}

\begin{abstract}
Observatory bibliographies increasingly need to determine not only whether a facility is named in a paper, but how its data support the scientific argument. We frame this need as semantic contribution attribution on a six-level scale (L0-L5), from non-use and mention-only cases to auxiliary, major, and exclusive data support. We present FacilityAudit-Agent, a framework that builds a page-bounded outline of each PDF, enriches sections with roles and facility mentions, and uses a tool-calling large language model (LLM) to fetch text on demand. Evidence from the sections is combined through an explicit decision tree and a coverage-calibrated confidence score, with uncertain cases routed to human review. The framework supports both single-facility and inventory audits and transfers to new facilities without retraining. On a test set of 39,358 human-annotated paper-facility pairs, FacilityAudit-Agent achieves 89.1\% accuracy and 79.8\% macro-F1 in \(\mathtt{single\_audit}\) mode, outperforming the other three baselines by 5.8, 8.3, and 12.1 accuracy points, respectively, with the clearest gains at L2 and L5. In \(\mathtt{inventory\_audit}\) mode, it also leads in facility listing and in level assignment among listed pairs. Ablations show multi-hop reading and tree indexing matter most for L3-L5. FacilityAudit-Agent has been deployed at the National Astronomical Data Center, where it is responsible for processing publications related to LAMOST, FAST, and EP.
\end{abstract}

\begin{keyword}
Astronomy data analysis, Astronomy databases, Computational astronomy, Facility bibliography, Large language models
\end{keyword}

\end{frontmatter}

\section{Introduction} \label{sec:intro}

Large-scale observational facilities are among the most expensive scientific infrastructures ever built, representing decades of engineering effort and substantial public investment \citep{harrison2023pathways}. Their operation, archive curation, funding allocation, and the planning of next-generation instruments all depend on a quantitative answer to one question: how much does each facility actually contribute to peer-reviewed papers \citep{d2023assessing}? In practice, this is measured by tracking how a facility's data are used in publications, which is the most direct indicator of scientific return \citep{grothkopf2011astronomy, winkelman2018indicators, dewitt2025insights}. At major observatories, this tracking is carried out by bibliography curation pipelines that must screen thousands of papers per year under tight budget and staffing constraints.

For decades, major observatories have manually compiled bibliographies of papers related to their facilities, a process that is cumbersome and time-consuming. Curators typically combine Astrophysics Data System (ADS) searches, author acknowledgments, program identifiers, and cross-checks against archive metadata \citep{grothkopf2003towards, chen2022best}. The European Southern Observatory (ESO) telescope bibliography (telbib), for example, requires librarians to verify that a paper actually analyzes ESO data rather than merely citing previous work, describing future observation plans, or discussing instrumentation \citep{grothkopf2018use}. Although manual curation is accurate, it is difficult to scale. With the growing volume of papers, especially from American Astronomical Society journals, many bibliographies lag behind by months or even years. More troublingly, a large number of project identifiers are not explicitly stated in the text and still require manual supplementation \citep{grothkopf2018use}.

To improve throughput, recent studies have applied machine learning (ML) and large language models (LLMs) to facility-related publication identification \citep{shao2024astronomical, amado2025identifying, wu2025amc}. A common practice is to frame this as a binary classification at the document level: if the facility name or any alias appears in the full text, the paper is counted \citep{chen2022classification}. We refer to the implicit assumption behind this strategy as ``mention-as-usage'': textual mention alone is taken as proof of scientific use, regardless of whether the study actually analyzed data from that facility \citep{d2023assessing}. This is computationally cheap and can achieve high recall when keywords are present. However, it conflates several qualitatively distinct types of citations (e.g., background citation, instrumental description, auxiliary calibration, and core data analysis) in the paper into a single positive label, and even fails to distinguish ambiguous content. This seriously obscures their substantive differences \citep{wu2025amc}.

The 2025 Telescope Reference and Astronomy Categorization Shared Task (TRACS) marked an important step toward finer-grained attribution by defining four categories: \textit{science}, \textit{instrumentation}, \textit{mention}, and \textit{not\_telescope} \citep{grezes2025overview}. Subsequently, Space Telescope Science Institute (STScI) introduced a new \textit{data-influenced} class \footnote{\url{https://www.stsci.edu/scientific-community/stsci-library-and-institutional-archive/missions-publications/bibliography-methodology}}. Nevertheless, these taxonomies remain relatively coarse. Even within the \textit{science} class, facilities vary in their substantive contributions to a paper's core claims: some facilities provide primary observational evidence, while others supply only supporting elements such as auxiliary calibration, cross-matching, or comparison benchmarks. Existing automated pipelines are largely unable to resolve this within-\textit{science} gradient, which is precisely the distinction that observatory bibliographers routinely make when deciding whether a paper belongs in a facility bibliography and how it should be tagged \citep{meakins2014two}.

These observations motivate us to reformulate the task. We argue that facility tagging should be understood as semantic contribution attribution: instead of asking only whether a facility is mentioned, the goal is to infer how strongly its data enter the paper's scientific argument. Doing so requires linking evidence across sections, rather than relying on isolated paragraphs or global keyword matching. To this end, we define a six-level contribution scale (L0-L5; Section~\ref{sec:description}) and propose FacilityAudit-Agent for fine-grained attribution in astrophysics papers. The framework builds a page-bounded table-of-contents (TOC) tree, enriches sections with roles and facility mentions, and uses a tool-calling language model to fetch text on demand. Evidence from data, analysis, and conclusion sections is then combined through an explicit decision tree, with supporting quotations and a confidence score for human review. The system supports both single-facility and inventory audits. We evaluate it against several LLM-based baselines on a human-annotated dataset of paper-facility pairs, and report ablation results, robustness across LLM backends, error analysis, and representative cases.

The paper is organized as follows. Section~\ref{sec:description} defines the attribution task and the L0-L5 scale. Section~\ref{sec:data} describes the dataset and annotation protocol. Section~\ref{sec:method} presents the FacilityAudit-Agent framework. Section~\ref{sec:experiments} details the evaluation metrics and experimental setup. Section~\ref{sec:results} reports the main results, ablations, cross-model tests, and remaining limitations. Section~\ref{sec:conclusion} concludes.

\section{Task Description} \label{sec:description}

\subsection{Semantic Contribution Attribution} \label{subsec:semantic}

Facility contribution attribution aims to determine, for each paper \(P\) and candidate facility \(f\), how \(f\) participates in the scientific argument of \(P\), rather than merely whether its name appears somewhere.

This distinction is crucial for observatory bibliographies: A paper may discuss the James Webb Space Telescope (JWST) in its introduction but analyze only archived data from HST; or it may use measurements from the Atacama Large Millimeter/submillimeter Array (ALMA) for calibration, while its main conclusions are based on observations from VLT. Treating such cases as equivalent ``facility papers'' obscures the actual scientific return of each observatory.

We formalize the task as follows. Let \(\mathcal{L}={\mathrm{L0},\mathrm{L1},\ldots,\mathrm{L5}}\) denote a six-level contribution scale (see Subsection~\ref{subsec:scale}), and let \(\mathcal{E}(P,f)\) be a set of textual evidence spans extracted from \(P\). The attribution function sought is \begin{equation} \mathcal{A}:(P,f)\mapsto \bigl(\ell,\mathcal{E}(P,f),\mathbf{s}\bigr),\quad \ell\in\mathcal{L}, \end{equation} where \(\mathbf{s}\) summarizes calibrated confidence and review flags derived from evidence coverage rather than from an unconstrained model self-score.

Reliable results of \(\mathcal{A}\) usually need evidence from more than one section. A facility name in the sample description may be insufficient to decide the level until the analysis and conclusions are also inspected. We refer to this cross-section linking as multi-hop reasoning. In contrast, classifying an isolated paragraph or counting global keyword hits often fails to distinguish background citation from actual data use.

Based on the daily needs of bibliographers, we provide two modes for this task.
Under \(\mathtt{single\_audit}\) mode, the target facility is given and the framework assigns a level, including L0 when the facility is absent or only ambiguously referenced. Under \(\mathtt{inventory\_audit}\) mode, the framework first lists observational facilities mentioned in the paper and then assigns a level to each listed facility.

\subsection{The Contribution Scale} \label{subsec:scale}

We operationalize contribution with six levels, L0-L5 (Table~\ref{tab:levels}). The scale is ordered by how strongly observational data from \(f\) enter the paper's scientific conclusions. L0 and L1 cover non-use and mention-only cases; L2 describes engineering or instrument-focused discussion without scientific analysis of \(f\)'s data; L3 marks auxiliary data use that does not enter the derivation of the main results; L4 and L5 distinguish major collaborative support from exclusive reliance on \(f\) for the core conclusions. Each level is defined by criteria that experts can check in the text: whether data from \(f\) are analyzed, whether conclusions depend on those data, and whether other observational sources are also required.

\begin{table*}[t]
\centering
\caption{Definition of contribution levels L0--L5 and correspondence to TRACS/STScI categories.}
\label{tab:levels}
\small
\setlength{\tabcolsep}{4pt}
\begin{tabular*}{\textwidth}{@{\hspace{0.5em}}c@{\extracolsep{\fill}}clc@{\hspace{0.5em}}}
\hline\hline
Level & Name & Description & TRACS/STScI \\
\hline
L0 & Non-relevant &
\parbox[t]{0.42\textwidth}{\raggedright \(f\) is not mentioned, or the acronym/string is ambiguous.} &
\parbox[t]{0.18\textwidth}{\centering \textit{not\_telescope}} \\[10pt]
L1 & Mention only &
\parbox[t]{0.42\textwidth}{\raggedright \(f\) appears in background or context (e.g., introduction) but no observational data from \(f\) are used in the analysis.} &
\parbox[t]{0.18\textwidth}{\centering \textit{mention}/\textit{data-influenced}} \\[14pt]
L2 & Instrument/engineering &
\parbox[t]{0.42\textwidth}{\raggedright The paper describes \(f\)'s design, calibration, or operations without analyzing \(f\)'s observational data for conclusions.} &
\parbox[t]{0.18\textwidth}{\centering \textit{instrumentation}/\textit{engineering}} \\[14pt]
L3 & Auxiliary contribution &
\parbox[t]{0.42\textwidth}{\raggedright Data from \(f\) support calibration, cross-matching, or comparison only; it is not used in deriving the main conclusions.} &
\parbox[t]{0.18\textwidth}{\centering \textit{science}} \\[14pt]
L4 & Major contribution &
\parbox[t]{0.42\textwidth}{\raggedright The main conclusions depend on data from \(f\), together with observational data from other facilities.} &
\parbox[t]{0.18\textwidth}{\centering \textit{science}} \\[14pt]
L5 & Exclusive contribution &
\parbox[t]{0.42\textwidth}{\raggedright All data supporting the core original conclusions come from \(f\); no other facilities were involved.} &
\parbox[t]{0.18\textwidth}{\centering \textit{science}} \\[14pt]
\hline
\end{tabular*}
\end{table*}

This scheme is finer than common bibliography tags but remains compatible with them. Table~\ref{tab:levels} provides the correspondence between this scheme and the categories used by TRACS and STScI. Those categories mainly decide whether a paper belongs in a facility bibliography; L0-L5 additionally decide how it belongs. In practice, a single science label may therefore correspond to L3, L4, or L5 depending on whether \(f\) provides auxiliary, major, or exclusive data support.

\section{Data} \label{sec:data}

\subsection{Data Acquisition and Annotation} \label{subsec:acquisition}

To construct a dataset suitable for L0-L5 attribution classification, we searched the NASA/ADS system for peer-reviewed papers published between 2022 and 2025 in 10 mainstream astronomical journals, including ApJ, ApJS and MNRAS. To minimise the occurrence of empty paper-facility pairs, we did not adopt random sampling; instead, we required that their full text or abstracts mention at least one of 10 astronomical facilities: LAMOST, VLT, Keck, GTC, Subaru, Gemini, EP, FAST, SDSS, and HST (see Appendix~\ref{app:abbreviation} for full names). After removing duplicates and non-peer-reviewed papers, we retained only those with machine-readable full-text PDFs (prioritizing arXiv, followed by publisher channels), ultimately obtaining 23,875 initial papers.

\begin{figure*}[th!]
	\centering 
	\includegraphics[width= 1 \textwidth]{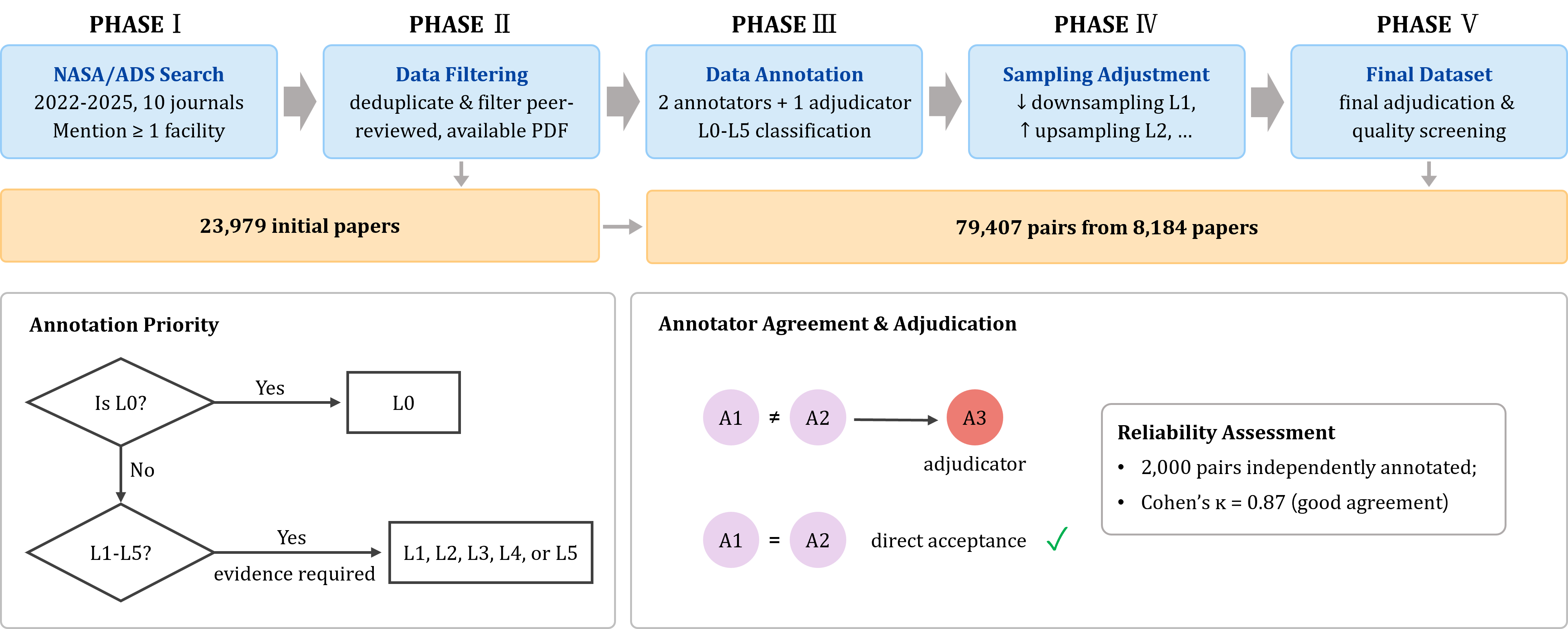}
	\caption{Overview of the data acquisition and annotation workflow.} 
	\label{fig:Acquisition_Annotation}
\end{figure*}

It is worth noting that, although the search criteria were limited to papers mentioning the above facilities, many L0 cases still appeared in the results. This was exactly as intended, because purely manual annotation rarely yields L0 cases. During the annotation phase, annotators were required to read the full text and first determine whether a pair belonged to L0; if not, they then assigned one of the L1-L5 levels. Annotation was conducted independently by two annotators, with a third annotator adjudicating discrepancies. A Cohen's \(\kappa\) of 0.87 (see Appendix~\ref{app:cohen}) was measured on a subset of 2,000 pairs, indicating good agreement. To more comprehensively evaluate the framework's performance, the dataset intentionally increased the sampling ratio of challenging cases, such as multi-facility joint analyses and mentions of facility names in acknowledgments. Moreover, while preserving the distribution of the real data, we performed downsampling on the large volume of L1 pairs and upsampling on the scarce L2 pairs as much as possible. Following final adjudication and quality screening, an annotated dataset comprising 79,407 paper-facility pairs from 8,184 different papers was established. The overall workflow of data acquisition and annotation is illustrated in Figure \ref{fig:Acquisition_Annotation}.

\subsection{Dataset Statistics and Split} \label{subsec:statistics}

The metadata for these final 79,407 pairs includes bibcode, facility name, and contribution level. The distribution of contribution levels is presented in Table~\ref{tab:level_dist}. L1 is the largest class, consistent with frequent background mentions of major facilities; L2 has the fewest samples, which aligns with the fact that pure instrument/engineering papers are inherently rare in the selected journals.

We split the dataset at the paper level into development, calibration, and test sets with approximate ratios of 15\% (1,227), 35\% (2,864), and 50\% (4,093), so that no paper appears in more than one split. This yields 11,886 development pairs, 28,163 calibration pairs, and 39,358 test pairs (see Table~\ref{tab:level_dist}). The development set is used only for prompt development and typical error analysis during the framework design; FacilityAudit-Agent does not involve fine-tuning based on these labels. The calibration set is reserved for temperature scaling and conformal threshold fitting of the confidence layer. The test set is used exclusively for final method comparison and evaluation.

\begin{table*}[t]
\centering
\caption{Contribution-level distribution of annotated paper-facility pairs and data splits}
\label{tab:level_dist}
\begin{tabular*}{\textwidth}{@{\hspace{1em}}c@{\extracolsep{\fill}}cccccc@{\hspace{1em}}}
\hline\hline
Level & Count & Percentage & Papers & Development Set & Calibration Set & Test Set \\
\hline
L0 & 2,992 & 3.8\% & 2,992 & 441 & 1,115 & 1,436 \\
L1 & 37,047 & 46.6\% & 7,125 & 5,664 & 13,184 & 18,199 \\
L2 & 668 & 0.8\% & 335 & 104 & 229 & 335 \\
L3 & 19,677 & 24.8\% & 5,390 & 2,829 & 6,956 & 9,892 \\
L4 & 15,534 & 19.6\% & 5,103 & 2,325 & 5,426 & 7,783 \\
L5 & 3,489 & 4.4\% & 2,930 & 523 & 1,253 & 1,713 \\
\hline
Total & 79,407 & 100\% & 23,875 (8,184) & 11,886 & 28,163 & 39,358 \\
\hline
\end{tabular*}
\\[2pt]
\footnotesize
\leftskip1em
\textit{Note.} The ``Papers'' column lists unique papers with at least one pair at that level; the column sum exceeds the corpus total because one paper may receive different levels for different facilities. The total unique paper count is 8,184.
\end{table*}

\section{Method} \label{sec:method}

\subsection{Overview of FacilityAudit-Agent} \label{subsec:overview}

\begin{figure*}[th!]
	\centering 
	\includegraphics[width= 1 \textwidth]{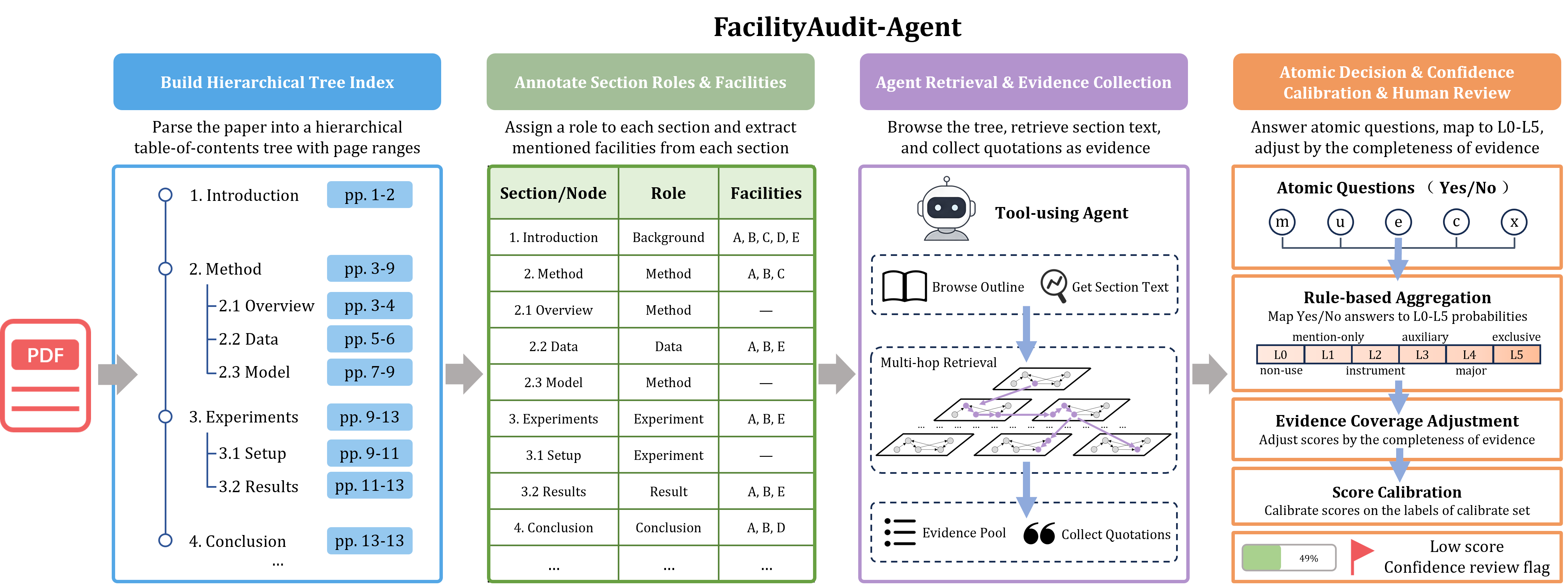}
	\caption{Overview of FacilityAudit-Agent (taking \(\mathtt{inventory\_audit}\) mode as an example)} 
	\label{fig:agent_pipeline}
\end{figure*}

FacilityAudit-Agent implements the attribution function \(\mathcal{A}\) defined in Subsection~\ref{subsec:semantic}. Given a paper \(P\) and a candidate facility, the framework returns a contribution level \(\ell\in\mathcal{L}\), a set of textual evidence spans \(\mathcal{E}(P,f)\), and a confidence summary \(\mathbf{s}\) for human review. When the facility is not specified in advance, the same pipeline can first list observational facilities mentioned in the paper and then audit each of them in turn. The framework is organized as a four-stage process (see Figure~\ref{fig:agent_pipeline}): (i) build a hierarchical tree outline of the paper; (ii) enrich outline nodes with section roles and facility mentions; (iii) use a tool-calling agent to gather and connect evidence across sections; and (iv) convert the agent's atomic assessments into a level and a review-oriented confidence score. Algorithm~\ref{alg:facility_audit} summarizes the end-to-end procedure; The remainder of this section (\ref{subsec:tree}-\ref{subsec:confidence}) details each stage in turn.

Three design choices guide the framework. First, retrieval is structured by the paper's own section hierarchy rather than by fixed-length text chunks or a vector index, as in retrieval-augmented generation (RAG). Astrophysics papers often place facility names, data descriptions, and scientific conclusions in different sections; keeping the table of contents makes it easier to move from a mention to the analysis that uses the data. Second, the agent fetches only the sections it needs. This avoids stuffing an entire long paper into a single context window and keeps the evidence trail inspectable. Third, the final level is not taken from an unconstrained model self-score; it is obtained from a small set of explicit atomic propositions that are combined by a fixed decision tree and then adjusted by checkable evidence-coverage factors.

\subsection{Hierarchical Tree Indexing} \label{subsec:tree}

The goal of the first stage is to reorganize paper \(P\) into a hierarchical outline, so that later stages can navigate effectively. The resulting index is denoted as
\begin{equation}
\mathcal{I}(P)=\bigl(T,\mathcal{P}\bigr),
\end{equation}
where \(T=(V,E)\) is a rooted tree whose nodes represent sections, and
\begin{equation}
\mathcal{P}=\bigl\{(i,\mathrm{text}_i)\bigr\}_{i=1}^{N}
\end{equation}
stores the full text page by page, with \(N\) being the total number of pages. Each node \(v\in V\) carries four pieces of information: a unique identifier, a section title, a one-sentence summary, and a contiguous page range
\begin{equation}
R(v)=[start_v, end_v],\ \ 1\le start_v\le end_v\le N.
\end{equation}
Child nodes correspond to subsections; leaf nodes are the units from which facility mentions are later extracted.

If the PDF contains a usable TOC, it is first parsed into a flat sequence of entries (each with a title and a starting page), then nested according to the numbering hierarchy (e.g., \(1\), \(1.1\), \(2.3.1\)) to form the tree. If no reliable TOC is available, candidate headings are instead detected from the body text, and a large language model (LLM) is asked to generate a reasonable section list constrained by the observed page order. In both cases, a section's end page is not simply set to one page before the next item in the flat TOC list. After the tree is assembled, the actual range of each section is recomputed based on sibling order and parent-child containment, with special care taken to avoid empty intervals when a parent and its first child start on the same page.

Concretely, suppose \(v^{1}, v^{2}, \ldots, v^{n}\) are the ordered children of the same parent (or the top-level sections if there is no parent), and let \(end_{\mathrm{limit}}\) denote the maximum allowed end page for this sibling group: for non-root nodes this is the parent's end page, and for root-level sections it is the total number of pages \(N\). Then the temporary end page of the \(i\)-th child, considering only sibling order, is determined as:
\begin{equation}
\label{eq:sibling-end}
end^{\mathrm{sib}}_{v^{i}} =
\begin{cases}
\max\bigl(start_{v^{i}},\, start_{v^{i+1}}-1\bigr), & i<n, \\[4pt]
\max\bigl(start_{v^{i}},\, end_{\mathrm{limit}}\bigr), & i=n.
\end{cases}
\end{equation}

If node \(v\) itself has children and its first child is \(c\), then the actual end page \(end_v\) is further tightened so that the parent's own text does not extend beyond the start page of its first child. This is done by taking:
\begin{equation}
\label{eq:parent-clamp}
end_v = \max\bigl(start_v,\,\min(start_c,\, end^{\mathrm{sib}}_v)\bigr).
\end{equation}
If \(v\) is a leaf node, \(end_v\) is simply set to \(end^{\mathrm{sib}}_v\). This correction is particularly useful for observational papers, where a major section heading and its first subsection heading often appear on the same page.

Additionally, each node may be supplemented with a short summary. For a leaf node, the summary is generated by extracting one sentence from its body text; for a parent node that has little or no text of its own, the summary is composed from the titles and summaries of its children. They are only for quick browsing and are not used in the final contribution-level decision. Moreover, this stage deliberately avoids fixed-length text chunking, because although chunking can retrieve locally similar passages, it breaks the logical chain from ``Data'' to ``Analysis'' to ``Conclusions'' that bibliographers follow when determining whether a facility was actually used.

\subsection{Structure Enrichment} \label{subsec:enrichment}

Tree indexing records only where each section is located. Structure enrichment adds two soft annotations to each node so that the audit agent can prioritize likely sections and jump to nodes that already name the candidate facility, instead of scanning the outline first. The first annotation is the role of the section (for example, introduction, observations, data analysis, results, or other), which serves as a soft prior over where contribution evidence is more likely to appear. The second is the set of facility names mentioned in that section, which seeds direct lookup of nodes that already name a given facility. Neither annotation assigns the contribution level. We denote the enriched tree by \(T^{+}\). For each node \(v\) in the tree node set \(V\), write \(M(v)\) for the facilities mentioned in its text.

Role assignment is performed automatically with simple title-matching rules. When a heading contains words such as ``Observation'', ``Data'', ``Analysis'', or ``Experiments and Results'', the corresponding role is assigned; remaining nodes are labeled other. If a heading is ambiguous, an LLM may optionally refine the assignment. Roles are soft hints, not hard filters: a node labeled other remains fully accessible later. In the agent's retrieval loop, the role only suggests which sections are more likely to contain usage statements and can therefore be examined earlier when the mention index alone is insufficient.

Facility mentions are extracted mainly on leaf nodes, where body text is densest. Candidate name strings in the text (for example, ``HST'' and ``Hubble Space Telescope'') are matched against a predefined list of facility names and aliases and then mapped to a canonical facility identifier. Mentions that clearly refer to an astronomical facility but are absent from the list are kept as open entries rather than discarded. All mentions are then aggregated into a paper-level facility index. For each facility \(f\), the index stores the nodes that mention \(f\):
\begin{equation}
\label{eq:facility-index}
\mathrm{Index}(f)=\bigl\{v\in V:\ f\in M(v)\bigr\}.
\end{equation} 
The page ranges covered by those nodes are recorded as well. When searching for evidence about \(f\), the agent can call a tool that returns \(\mathrm{Index}(f)\) and then fetch only those nodes; affiliated instrument or survey names linked to \(f\) in the alias list may also be included, but they do not by themselves determine the contribution level of \(f\). The output of this stage is the pair \((T^{+},\mathrm{Index})\), which later tools use when listing facilities or locating seed evidence.

\subsection{Facility Audit Agent} \label{subsec:agent}
The audit stage is a tool-using loop bound to one paper-facility pair. Let \(h_t\) be the dialogue history after round \(t\), and let the tool set be
\begin{equation}
\Theta=
\left\{
\begin{aligned}
&\mathtt{inspect\_doc},\,
\mathtt{browse\_outline},\\
&\mathtt{expand\_node},\,
\mathtt{fetch\_section},\\
&\mathtt{fetch\_content},\,
\mathtt{find\_evidence},\\
&\mathtt{list\_facilities}
\end{aligned}.
\right.
\end{equation}
At round \(t\le Round_{\max}\), the framework selects an action \(a_t\in\Theta\) and arguments \(\theta_t\), runs the tool against the index \(\mathcal{I}(P)\) and the enriched tree \(T^{+}\), and receives the tool output \(o_t\). The evidence set is updated as follows:
\begin{equation}
\mathcal{E}_{t}=\mathcal{E}_{t-1}\ \cup \ \mathrm{spans}(o_t),\qquad
\mathcal{E}_{0}=\emptyset,
\end{equation}
while the history is compacted according to
\begin{equation}
h_t=\operatorname{Compact}(h_{t-1},a_t,\theta_t,o_t),
\end{equation}
so that long lists generated early in the process do not crowd out the more recently fetched section texts. The loop terminates early as soon as the agent emits a valid final judgment or when the evidence and history \((\mathcal{E}, h)\) are deemed adequate for a decision.

The order in which the agent reads sections is fixed and deliberately conservative. It first calls \(\mathtt{inspect\_doc}\) to verify the paper's status and total page count. Next, it calls \(\mathtt{find\_evidence}\) without any role filter; this tool returns all nodes that have been linked to facility \(f\) (and optionally its related names) via Equation~\eqref{eq:facility-index}. Those hit nodes are then read using \(\mathtt{fetch\_section}\). Only when the hit nodes are insufficient or missing does the agent expand the outline by titles, inspect sibling or parent sections using \(\mathtt{browse\_outline}\) or \(\mathtt{expand\_node}\), or directly retrieve specific page ranges via \(\mathtt{fetch\_content}\). This reading order matches the multi-hop pattern described in Section~\ref{subsec:semantic}. The agent is asked to keep short verbatim quotations, and when the level assignment depends on multiple quotation spans, the reasoning must explicitly state the connections among those spans. The audit loop above is unchanged in both operating modes.

The agent's final output is a JSON object containing the operating mode (\(\mathtt{single}\) or \(\mathtt{inventory}\)), the facility name, a proposed level \(\ell_{\mathrm{agent}}\), supporting quotations, and the five atomic assessments. In \(\mathtt{single\_audit}\) mode, \(\ell_{\mathrm{agent}}\) may be any label in \(\{\mathrm{L0},\ldots,\mathrm{L5}\}\). In \(\mathtt{inventory\_audit}\) mode, only listed observational facilities are returned, and each reported level lies in \(\{\mathrm{L1},\ldots,\mathrm{L5}\}\); software packages, pipelines, unrelated project names, etc., are omitted from the inventory rather than forced to L0. In other words, a true L0 facility should not appear in the predicted inventory. Moreover, any output that violates the JSON schema, invents quotations not present in the tool results, or assigns levels to ineligible items is rejected by the guardrails.

\subsection{Atomic Assessments and Level Logic} \label{subsec:atomic}

To keep levels auditable, FacilityAudit-Agent decomposes each paper-facility decision into five atomic propositions after collecting the evidence:
\begin{itemize}
\item \(m\): does the text establish the intended facility (rather than a homonym or unrelated acronym)?
\item \(u\): given \(m\), are observational data from \(f\) used in the paper?
\item \(e\): given \(m\) and \(\neg u\), does the paper contain substantive engineering or instrument discussion of \(f\)?
\item \(c\): given \(m\) and \(u\), do the core original conclusions depend on data from \(f\)?
\item \(x\): given \(m\), \(u\), and \(c\), do all observational data supporting those conclusions come exclusively from \(f\)?
\end{itemize}
Each answer is reported as a discrete status
\begin{equation}
\sigma\in\{\mathtt{supported},\mathtt{contradicted},\mathtt{ambiguous},\mathtt{unknown}\}.
\end{equation}
These assessments must be turned into soft scores in \([0,1]\) before they can enter the decision tree. We use fixed cold-start priors that are then pulled toward \(1/2\) according to the completeness of the assessment. For facility identity, an additional score \(m_{\mathrm{scan}}\) is obtained by a direct alias scan over the full text. The fused identity probability is:
\begin{equation}
\label{eq:identity}
m=\lambda\, m_{\mathrm{scan}}+(1-\lambda)\, m_{\mathrm{model}},\qquad \lambda=0.7,
\end{equation}
so that an ambiguous abbreviation cannot receive a high \(m\) from string matching alone. Denoting the mapped probabilities of the remaining questions as \(u,e,c,x\), the six contribution levels follow a mutually exclusive decision tree:
\begin{align}\label{eq:tree}
\Pr(\mathrm{L0}\mid P,f) &= 1-m, \nonumber\\
\Pr(\mathrm{L1}\mid P,f) &= m(1-u)(1-e), \nonumber\\
\Pr(\mathrm{L2}\mid P,f) &= m(1-u)e, \nonumber\\
\Pr(\mathrm{L3}\mid P,f) &= mu(1-c), \nonumber\\
\Pr(\mathrm{L4}\mid P,f) &= muc(1-x), \nonumber\\
\Pr(\mathrm{L5}\mid P,f) &= mucx.
\end{align}

By construction, these six probabilities sum to one. This decision tree serves as the operational implementation of Table~\ref{tab:levels}, and incorporates the discriminative logic adopted by human annotators. Each branch of the tree maps to a clear judgment condition. As a result, bibliographers can not only view the final assigned level but also trace which intermediate judgment leads to that level. This decoupling design is intentional. A direct, general request about ``contribution level'' tends to yield responses that are fluent but poorly supported by evidence. In contrast, requesting the explicit output of the variables \(m,u,e,c,x\) forces the level label to be tightly coupled with each verification condition, thereby facilitating cross-checking against the quoted evidence.

\subsection{Confidence, Calibration, and Review} \label{subsec:confidence}

Level probabilities of Equation~\eqref{eq:tree} assume that the supporting evidence has been adequately inspected. For brevity, we write
\begin{equation}
p=\bigl(p_{\mathrm{L0}},\ldots,p_{\mathrm{L5}}\bigr),\qquad
p_{\mathrm{L}k}=\Pr(\mathrm{L}k\mid P,f),
\end{equation}
for the six level probabilities generated by that tree. In practice, retrieval may miss a key section, a quotation may fail verification, or the alias scan may be incomplete. FacilityAudit-Agent therefore multiplies each class probability by an evidence-coverage factor \(\gamma_k\in(0,1]\):
\begin{equation}\label{eq:coverage}
C_k=p_{\mathrm{Lk}}\,\gamma_k,\qquad
k=0,\ldots,5.
\end{equation}
Each \(\gamma_k\) is a weighted geometric mean of coverage signals appropriate to that level, including full-text availability, alias-scan coverage, quotation grounding against the fetched text, relevant-section coverage, completeness of the multi-hop conclusion chain, and, for L5, whether competing facilities in the inventory have been considered. A geometric mean penalizes any critical factor near 0. The residual
\begin{equation}
U=1-\sum_{k=0}^{5}C_k
\end{equation}
is treated as unknown probability reserved for incomplete evidence. The scalar confidence reported for the chosen level is \(C_{\ell}\) (or the coverage-adjusted score of the recommended level after calibration). Without an independent calibration fit, this quantity is an evidence score suitable for ranking and triage; it should not be read as a fully calibrated probability.

In addition, a temperature \(\tau>0\) is chosen on the calibration set of Section~\ref{sec:data} so that, after the rescaling
\begin{equation}
\label{eq:temperature}
\tilde{p}_{\mathrm{L}k}=\frac{p_{\mathrm{L}k}^{1/\tau}}{\sum_{j\in\mathcal{L}} p_{\mathrm{L}j}^{1/\tau}},
\end{equation}
the predicted levels match the human-annotated levels as well as possible on that set. In short, \(\tau\) stretches or compresses the six scores so that a reported value is closer to how often that level is actually correct. Temperature scaling affects only \(\mathbf{s}\) and the review trigger; it does not change the decision tree or the cold-start mapping.

A conformal threshold can further return a prediction set that is designed to contain the true level for at least 90\% (our default) of cases when the calibration pairs are drawn from the same distribution as the cases being scored. At deployment time, a case is flagged for human review when any of the following holds: the confidence is low; the agent level and the coverage-adjusted recommendation disagree; the unknown probability is high; the prediction set is large; or calibration parameters are unavailable. High-confidence, singleton predictions can pass with lighter inspection, while flagged cases enter the bibliographer queue. In this sense, FacilityAudit-Agent is designed not to replace human judgment, but as a scalable screening layer ahead of bibliographer curation.

\begin{algorithm}[t]
\caption{FacilityAudit-Agent}
\label{alg:facility_audit}
\begin{algorithmic}[1]
\REQUIRE Paper $P$; mode $\in\{\mathtt{single},\mathtt{inventory}\}$; optional candidate $f^{\star}$; max rounds $Round_{\max}$
\ENSURE Set of records $\{(\ell,\mathcal{E},\mathbf{s})\}$ for audited facilities
\STATE $\mathcal{I}(P)\leftarrow\textsc{BuildTree}(P)$ \COMMENT{Subsection~\ref{subsec:tree}}
\STATE $T^{+}\leftarrow\textsc{Enrich}(\mathcal{I}(P))$ \COMMENT{Subsection~\ref{subsec:enrichment}}
\IF{$\mathrm{mode}=\mathtt{single}$}
\STATE $\mathcal{F}\leftarrow\{f^{\star}\}$
\ELSE
\STATE $\mathcal{F}\leftarrow\textsc{Filter}(\mathtt{list\_facilities}(T^{+}))$
\ENDIF
\FOR{each facility $f\in\mathcal{F}$}
\STATE $\mathcal{E}\leftarrow\emptyset$; $h\leftarrow\emptyset$
\FOR{$t=1$ to $Round_{\max}$}
\STATE $(a_t,\theta_t)\leftarrow\textsc{AgentPolicy}(h,T^{+},f)$
\STATE $o_t\leftarrow\textsc{ExecTool}(a_t,\theta_t;\mathcal{I}(P),T^{+})$
\STATE $\mathcal{E}\leftarrow\mathcal{E}\cup\mathrm{spans}(o_t)$
\STATE $h\leftarrow\textsc{Compact}(h,a_t,\theta_t,o_t)$
\IF{$\textsc{ReadyToDecide}(\mathcal{E},h)$}
\STATE \textbf{break}
\ENDIF
\ENDFOR
\STATE $(\sigma_m,\sigma_u,\sigma_e,\sigma_c,\sigma_x),\ell_{\mathrm{agent}}\leftarrow\textsc{AtomicAssess}(\mathcal{E},f)$
\STATE $(m,u,e,c,x)\leftarrow\textsc{MapPriors}(\sigma_{\cdot},m_{\mathrm{scan}})$
\STATE $p\leftarrow\textsc{DecisionTree}(m,u,e,c,x)$ \COMMENT{Equation~\eqref{eq:tree}}
\STATE $(C,U,\mathbf{s})\leftarrow\textsc{Confidence}(p,\mathcal{E},T^{+})$ \COMMENT{Equation~\eqref{eq:coverage}}
\STATE $\ell\leftarrow\textsc{SelectLevel}(C,\ell_{\mathrm{agent}},\mathbf{s})$
\STATE \textbf{output} $(\ell,\mathcal{E},\mathbf{s})$
\ENDFOR
\end{algorithmic}
\end{algorithm}

\section{Experiments} \label{sec:experiments}

\subsection{Evaluation Metrics} \label{subsec:metrics}

All experiments use the test split in Section~\ref{sec:data}. We report two evaluation settings that match the two modes.

\subsubsection{Single-facility level attribution} \label{subsec:metrics_single}

In \(\mathtt{single\_audit}\) mode, each baseline receives a paper-facility pair \((P,f)\) with a fixed target facility and predicts a level \(\hat{\ell}\in\mathcal{L}\). The prediction is compared with the human-annotated label \(\ell\). We report overall accuracy,
\begin{equation}
\mathrm{Accuracy}=\frac{1}{n}\sum_{i=1}^{n}
\begin{cases}
1, & \hat{\ell}_i=\ell_i, \\
0, & \hat{\ell}_i\neq\ell_i,
\end{cases}
\end{equation}
where \(n\) is the number of test pairs. Because frequent levels such as L1 can dominate accuracy, we also compute precision, recall, and F1 for each level. Writing \(\mathrm{TP}_k\), \(\mathrm{FP}_k\), and \(\mathrm{FN}_k\) for the corresponding counts,
\begin{gather}
\mathrm{Precision}_k=\frac{\mathrm{TP}_k}{\mathrm{TP}_k+\mathrm{FP}_k}, \notag\\[4pt]
\mathrm{Recall}_k=\frac{\mathrm{TP}_k}{\mathrm{TP}_k+\mathrm{FN}_k}, \notag\\[4pt]
\mathrm{F1}_k=\frac{2\,\mathrm{Precision}_k\cdot\mathrm{Recall}_k}{\mathrm{Precision}_k+\mathrm{Recall}_k}.
\end{gather}
Macro-averaged precision, recall, and F1 are
\begin{equation}
\mathrm{Macro\text{-}P}=\frac{1}{|\mathcal{L}|}\sum_{k\in\mathcal{L}}\mathrm{Precision}_k,
\end{equation}
and likewise for Macro-R and Macro-F1, so that rare levels such as L2 are not hidden.
We additionally report per-level support and a \(6\times6\) confusion matrix.

\subsubsection{Inventory listing and level attribution} \label{subsec:metrics_inventory}

In \(\mathtt{inventory\_audit}\) mode, each baseline receives only the paper, generates a facility list, and assigns a level to each listed observational facility.
Inventory mode does not output L0; listed facilities receive labels in \(\{\mathrm{L1},\ldots,\mathrm{L5}\}\) only (Subsection~\ref{subsec:agent}). Evaluation therefore has two parts: whether the right facilities are listed, and whether the assigned levels are correct. For all baselines, the facility list itself is first extracted by an LLM pass over the paper or over the method's retrieved snippets.

\emph{Facility listing.} Consider every test pair whose human label is L1-L5 as a pair that should appear in the inventory, and every pair returned by the method as a predicted listing. Listing counts are defined in the usual way: \(\mathrm{TP}_{\mathrm{list}}\) is a predicted listing whose facility matches an annotated L1-L5 pair on the same paper; \(\mathrm{FP}_{\mathrm{list}}\) is a predicted listing with no matching L1-L5 annotation (including cases where the annotated label is L0); and \(\mathrm{FN}_{\mathrm{list}}\) is an annotated L1-L5 pair absent from the predicted inventory. Listing precision, recall, and F1 follow the same formulas as above, with subscript \(\mathrm{list}\) in place of a level index \(k\).

\emph{Level attribution on matched pairs.} For level scoring, we keep only annotated pairs with labels in \(\{\mathrm{L1},\ldots,\mathrm{L5}\}\) that also appear in the predicted inventory. On this matched subset, we compute each evaluation metric, and a \(5\times5\) confusion matrix, using the same equations as above but with \(|\mathcal{L}|=5\). Annotated L0 pairs are excluded from the level matrix.

\subsection{Experimental Setup} \label{subsec:setup}

In practical facility tagging, although keyword matching and traditional ML methods remain common, they are not suitable for the task in this paper: the decision depends on how a named facility is used in context, not on whether a name string occurs. We therefore compare only baselines below based on LLMs, which differ mainly in how they read the paper.

\textbf{Full-context-LLM} concatenates the paper's full text with the L0-L5 level definitions into a single long prompt, feeds it into an LLM, and directly asks a level prediction. When the combined prompt exceeds the model's context window, the body is truncated from the end until it fits, while the title, abstract, and the level definitions are kept. This baseline serves to evaluate the performance ceiling achievable by pure long-context question answering, without any external tools or explicit decision trees.

\textbf{TRACS-AMC} follows the Automated Mission Classifier (AMC) of \cite{wu2025amc}.
Given a paper and a candidate facility, it first keeps sentences that match facility-related keywords and expands each hit into a local window of 3 sentences on either side. Those snippets are reranked by a lightweight LLM for relevance to facility usage, and the top 15 snippets are passed to a second LLM that predicts the four TRACS categories. Since the original AMC's \textit{science} category cannot distinguish L3, L4, and L5 in a fine-grained manner, we made a minimal adaptation to match the attribution task studied here. While keeping the AMC evidence pipeline unchanged, we replaced the four-level output head with a six-level decision over L0-L5 (Table~\ref{tab:levels}), and let the second LLM judge the extent to which the facility's data contribute to the paper's conclusions.

\textbf{Chunk-RAG} adopts a standard RAG architecture that integrates a vector database and semantic similarity search. This contrasts with TRACS-AMC, which prioritizes flexibility and cost-effectiveness. It splits the paper into fixed-length chunks of 1024 tokens with an overlap of 128 tokens, embeds each chunk in a vector index, ranks chunks by semantic similarity to a query built from the name and aliases of the candidate facility, keeps the top \(k=5\) chunks, and concatenates them in rank order until the context budget is filled before asking the LLM for a level. This baseline represents the common retrieve-then-classify pipeline and serves as the closest contrast to our framework of preserving the section tree rather than breaking the paper into unrelated windows.

\textbf{FacilityAudit-Agent} uses the four-stage pipeline in Section~\ref{sec:method}: tree indexing, structure enrichment, tool-based multi-hop reading, and coverage-adjusted level selection. To measure what each stage contributes, we also evaluate four ablated variants (``w/o'' = without), each with one component removed.
(i) \textbf{w/o tree}: the outline is discarded and the agent may only fetch page spans, as in ordinary full-text paging.
(ii) \textbf{w/o enrichment}: section roles and the facility index in Equation~\eqref{eq:facility-index} are not built, so the agent must locate mentions from the bare outline.
(iii) \textbf{w/o multi-hop}: the agent is not instructed to follow evidence from sample description to analysis to conclusions, and may stop after the first hit section.
(iv) \textbf{w/o SelectLevel}: the reported label is the agent's raw $\ell_{\mathrm{agent}}$, without coverage-adjusted scores or the review flag.
Prompts, alias lists, and these design choices are fixed using only the development split.

The main comparison and all ablations use the same LLM, DeepSeek-V4-Flash. Under this LLM, Full-context-LLM, TRACS-AMC, Chunk-RAG, FacilityAudit-Agent, and the four ablated variants share the same generation settings, so that differences among baselines arise primarily from how they read the paper and determine the level, rather than from changes in model or sampling parameters. To verify that the results do not depend on a single LLM API, we rerun FacilityAudit-Agent with four additional models: GPT-5.5, GLM-5.2, Qwen3.6-27B, and Kimi K2.6. For FacilityAudit-Agent, the tool loop is capped at \(Round_{\max}=30\). Each ablated variant retains this cap and, where still applicable, uses the same tools. Temperature \(\tau\) and the conformal threshold, when used, are taken from the calibration procedure in Subsection~\ref{subsec:confidence}.

\section{Results and Discussion} \label{sec:results}

\subsection{Single-facility comparison}
\label{subsec:main_single}

Table~\ref{tab:single_results} summarizes the comparison results for the \texttt{single\_audit} mode on the test set. FacilityAudit-Agent achieves an accuracy of 89.1\% and a macro-average F1 of 79.8\%, outperforming Full-context-LLM by 5.8 and 8.7 percentage points, respectively, and surpassing Chunk-RAG and TRACS-AMC by still larger margins. The gain is most visible where mention-as-usage and local-snippet strategies typically fail: L2 and L5 F1 rise substantially relative to Full-context-LLM and Chunk-RAG, and L0 discrimination also improves. These levels require verifying exclusive or instrument-focused use against competing facilities across analysis and conclusion sections. Full-context-LLM remains competitive on L3-L4, but its single-pass reading yields lower L5 precision and more L1\(\to\)L0 confusions than FacilityAudit-Agent. Chunk-RAG keeps a high L1 F1, yet frequently demotes true L1 pairs to L0 and remains weak on L5, consistent with assigning levels from a few top-ranked fixed-length chunks retrieved by semantic similarity rather than following the evidence chain across sections. TRACS-AMC, even after mapping its output to L0-L5, is weakest on L2 and L5: short keyword windows recover mentions efficiently but provide little support for separating auxiliary, major, and exclusive data use.

\begin{table*}[th!]
\centering
\caption{Single-facility comparison (\texttt{single\_audit} mode) results on the test set (n=39,358). Left: per-level F1. Right: overall accuracy and macro-averaged precision, recall, and F1. All entries are in percent (\%).}
\label{tab:single_results}
\small
\setlength{\tabcolsep}{5pt}
\begin{tabular*}{\textwidth}{@{\hspace{1em}}l@{\extracolsep{\fill}}cccccccccc@{\hspace{1em}}}
\hline\hline
& \multicolumn{6}{c}{Per-level F1} & \multicolumn{4}{c}{Overall} \\
\cmidrule(l{-1pt}r{4pt}){2-7} \cmidrule(r{0pt}){8-11}
Method & L0 & L1 & L2 & L3 & L4 & L5 & Acc & Macro-P & Macro-R & Macro-F1 \\
\midrule
Full-context-LLM     & 74.5 & 91.6 & 48.5 & 82.5 & 79.6 & 50.0 & 83.3 & 67.0 & 79.1 & 71.1 \\
TRACS-AMC            & 71.6 & 88.7 & 38.5 & 74.6 & 71.1 & 36.1 & 77.0 & 59.8 & 71.5 & 63.4 \\
Chunk-RAG            & 68.5 & 91.6 & 45.9 & 80.0 & 74.7 & 40.7 & 80.8 & 63.0 & 74.8 & 66.9 \\
\textbf{FacilityAudit-Agent} & \textbf{83.8} & \textbf{95.5} & \textbf{62.3} & \textbf{87.8} & \textbf{85.5} & \textbf{63.9} & \textbf{89.1} & \textbf{75.9} & \textbf{85.7} & \textbf{79.8} \\
\hline
\end{tabular*}
\end{table*}

\begin{figure*}[th!]
    \centering
    \includegraphics[width= 1 \textwidth]{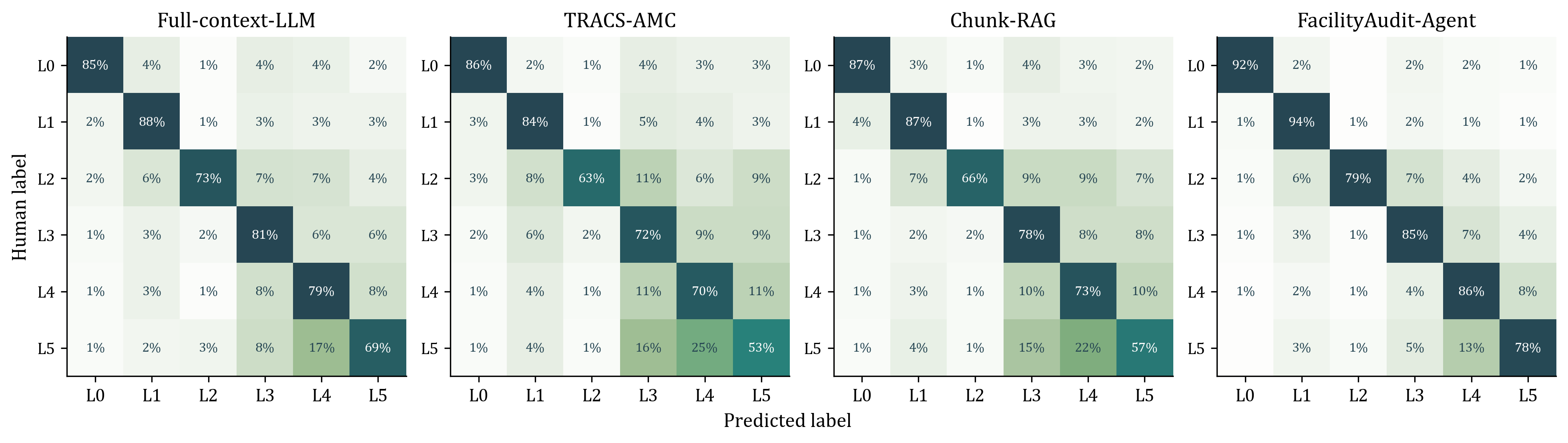}
    \caption{Row-normalized \(6\times6\) confusion matrices under \texttt{single\_audit} mode.}
    \label{fig:single_matrix}
\end{figure*}

The confusion matrices in Figure~\ref{fig:single_matrix} clarify where each method still errs. For FacilityAudit-Agent, off-diagonal mass concentrates on neighboring levels (especially L3\(\leftrightarrow\)L4 and L4\(\leftrightarrow\)L5) rather than long-range swaps. Despite L1 dominating the test set (46.2\% of pairs), recall on L3 and L4 stays comparatively strong, whereas L2 and L5 remain harder given their small counts. Full-context-LLM spreads errors more widely among L3-L5; Chunk-RAG and TRACS-AMC over-assign L5 for many L3/L4 cases and also send many L1 pairs to L0. Overall, preserving the section tree and linking evidence across observations, analysis, and conclusions improves rare-class recall and boundary discrimination beyond flat long-context prompting or snippet retrieval alone.

\subsection{Inventory comparison}
\label{subsec:main_inventory}

Table~\ref{tab:inventory_results} shows the \texttt{inventory\_audit} results. Because each method must first list observational facilities and then assign a level to every listed entry, the evaluation separates facility listing from level attribution on the matched L1-L5 subset. FacilityAudit-Agent leads both stages, with 83.3\% listing F1 and 78.6\% macro-F1 on matched pairs. The separation among baselines arises mainly from how much paper text each method exposes to the listing step. Full-context-LLM conditions extraction on the longest available context (the full paper, truncated only when necessary), which tends to suppress spurious mentions and yields higher precision but lower recall (59.2\%). Chunk-RAG and TRACS-AMC rely on shorter, locally retrieved spans, so a facility is listed whenever its name or alias appears in a retrieved segment; repeated mentions across segments can surface the same facility multiple times, which raises recall (about 80\%-82\%) but also introduces unrelated entities into the inventory and lowers precision. FacilityAudit-Agent draws on the enriched outline and facility index rather than a single long flat prompt or isolated snippets, and thereby keeps both listing precision and recall high.

\begin{table*}[th!]
\centering
\caption{Inventory comparison (\texttt{inventory\_audit} mode) results on the test set. Left: facility-listing precision, recall, and F1 over annotated L1-L5 pairs (n=37,922 should-list). Right: per-level F1 on the matched L1-L5 subset and matched macro-F1. All entries are in percent (\%).}
\label{tab:inventory_results}
\small
\setlength{\tabcolsep}{4.5pt}
\begin{tabular*}{\textwidth}{@{\hspace{1em}}l@{\extracolsep{\fill}}cccccccccc@{\hspace{1em}}}
\hline\hline
& \multicolumn{3}{c}{Facility listing} & \multicolumn{6}{c}{Matched level attribution} \\
\cmidrule(l{1pt}r{3pt}){2-4} \cmidrule(l{1pt}r{-16pt}){5-10}
Method & P & R & F1 & L1 & L2 & L3 & L4 & L5 & Macro-F1 \\
\midrule
Full-context-LLM     & 80.5 & 59.2 & 68.2 & 92.2 & 46.8 & 81.6 & 78.9 & 49.4 & 69.8 \\
TRACS-AMC            & 60.0 & \textbf{81.8} & 69.2 & 89.6 & 40.1 & 74.8 & 69.4 & 34.9 & 61.8 \\
Chunk-RAG            & 62.0 & 79.7 & 69.7 & 93.8 & 49.9 & 77.5 & 70.4 & 37.2 & 65.7 \\
\textbf{FacilityAudit-Agent} & \textbf{86.5} & 80.4 & \textbf{83.3} & \textbf{96.2} & \textbf{59.2} & \textbf{88.3} & \textbf{86.1} & \textbf{63.3} & \textbf{78.6} \\
\hline
\end{tabular*}
\end{table*}

On the matched subset, FacilityAudit-Agent maintains the largest margin over the baselines in macro-F1, with the clearest gains at L3-L5. Full-context-LLM remains competitive on L3-L4, but its lower listing recall shrinks the matched evaluation pool and leaves many annotated pairs ungraded. Chunk-RAG and TRACS-AMC retain strong matched L1 scores yet drop sharply on L5, consistent with level assignment from short snippets after a high-recall listing step. L2 remains the weakest class for all methods, reflecting both its scarcity and errors that begin at listing, where true L2 pairs may be omitted or listed without the instrument-focused sections needed for correct labeling.

\begin{figure*}[th!]
    \centering
    \includegraphics[width=\textwidth]{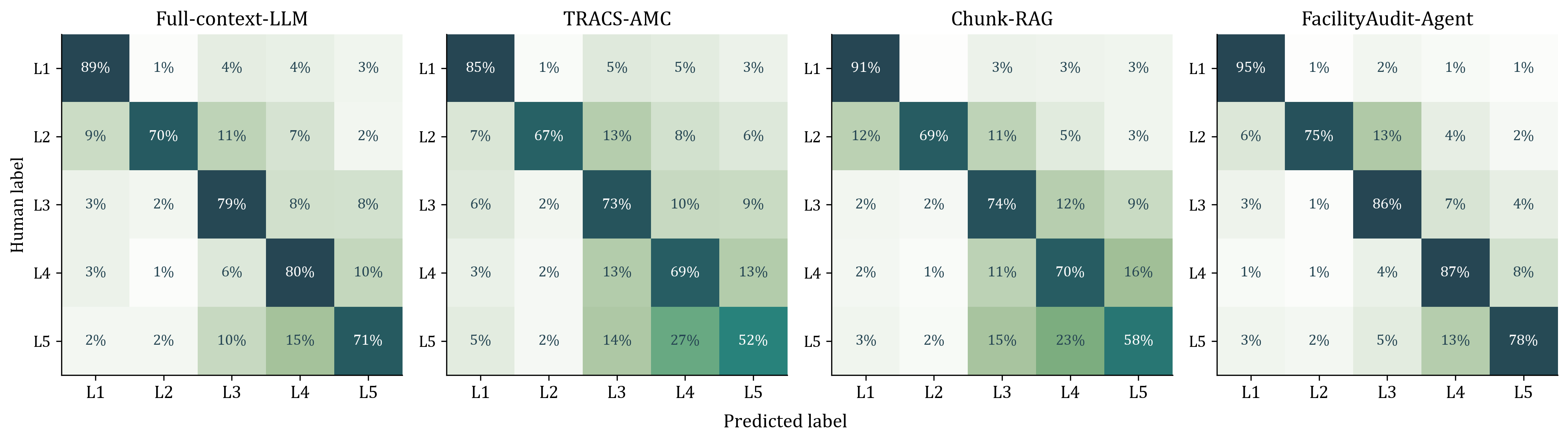}
    \caption{Row-normalized \(5\times5\) confusion matrices for matched level attribution under \texttt{inventory\_audit} mode.}
    \label{fig:inventory_matrix}
\end{figure*}

Figure~\ref{fig:inventory_matrix} presents the \(5\times5\) confusion matrices on the matched subset. FacilityAudit-Agent's misclassifications are still concentrated among adjacent levels (especially L3\(\leftrightarrow\)L4 and L4\(\leftrightarrow\)L5) rather than long-range swaps. Full-context-LLM has a relatively compact error distribution but evaluates fewer samples overall. Chunk-RAG and TRACS-AMC show more severe L3/L4\(\to\)L5 confusion after their high-recall listing step. Overall, inventory-style attribution requires both sufficient listing context to control false positives and structured reading to resolve L3-L5 boundaries once facilities have been listed; neither a conservative full-text approach nor a snippet-level high-recall strategy alone is sufficient for this task.

\subsection{Ablation study}
\label{subsec:ablation}

\begin{figure*}[th!]
    \centering
    \includegraphics[width=\textwidth]{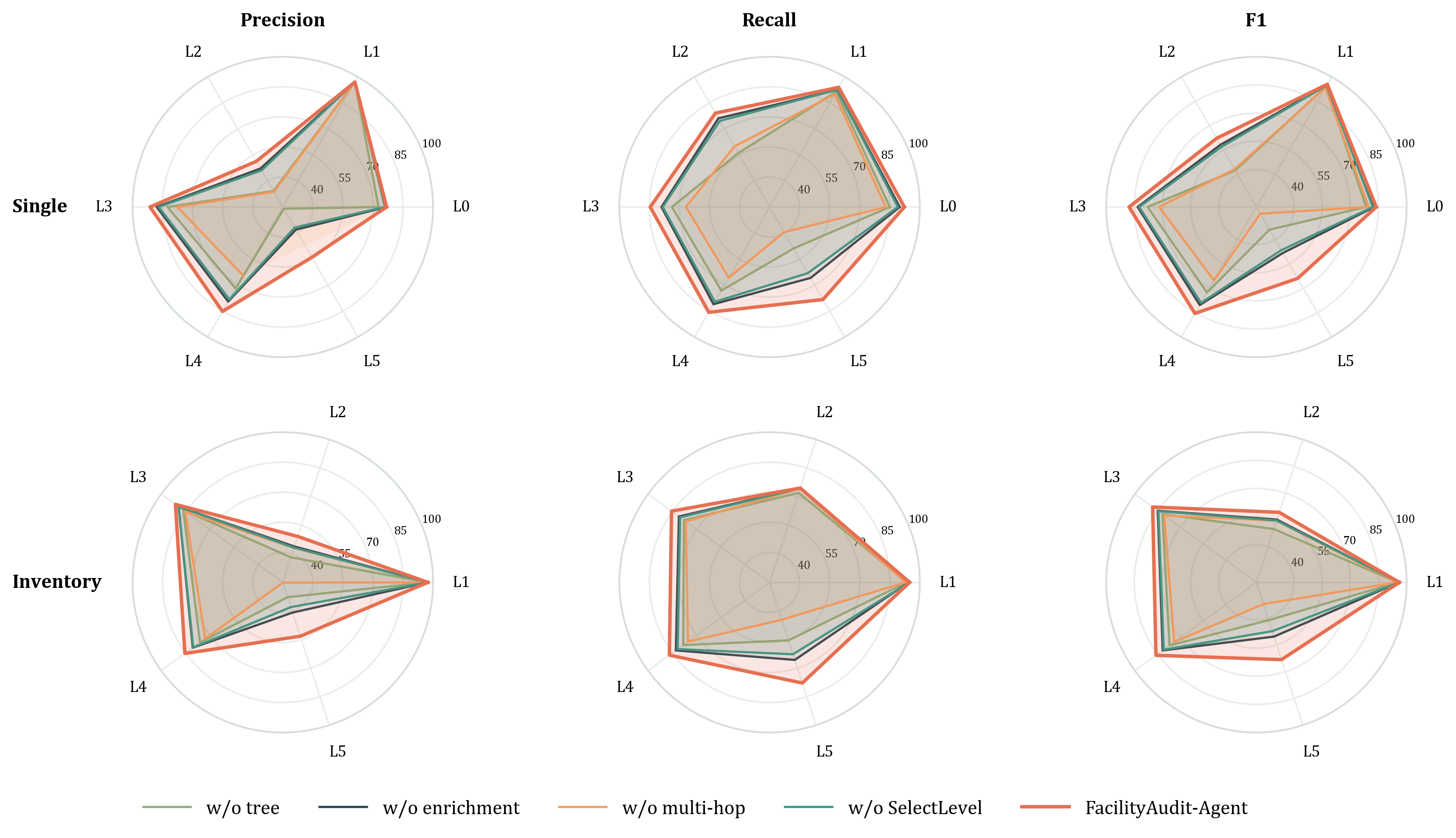}
    \caption{Per-level ablation profiles under \texttt{single\_audit} (top; L0-L5) and \texttt{inventory\_audit} (bottom; matched subset, L1-L5). Each panel shows one metric (precision, recall, or F1); each polygon corresponds to one ablated variant. FacilityAudit-Agent is drawn with a heavier line.}
    \label{fig:ablation_radar}
\end{figure*}

Table~\ref{tab:ablation_results} and Figure~\ref{fig:ablation_radar} present the ablation study of FacilityAudit-Agent under both \texttt{single\_audit} and \texttt{inventory\_audit}. The former summarizes accuracy and macro-averaged precision, recall, and F1; the latter shows each variant by level and metric.

Under \texttt{single\_audit} (n=39,358; L0-L5) mode, removing cross-section multi-hop reading (\textbf{w/o multi-hop}) causes the largest drop (76.7\% accuracy, 62.9\% macro-F1), followed by \textbf{w/o tree}, \textbf{w/o SelectLevel}, and \textbf{w/o enrichment}. Macro-F1 falls by 16.9, 13.2, 6.2, and 5.2 percentage points, respectively. In Figure~\ref{fig:ablation_radar} (top row), the \textbf{w/o multi-hop} polygon collapses most at L3-L5, especially in the F1 panel, whereas L1 spokes remain close to the full system across variants.

Under \texttt{inventory\_audit} mode, metrics in Table~\ref{tab:ablation_results} are computed on the matched listed subset (L1-L5). The full framework reaches 89.8\% accuracy and 78.6\% macro-F1 on 30,489 matched pairs. Removing multi-hop again yields the largest matched macro-F1 drop, followed by tree, SelectLevel, and enrichment. Figure~\ref{fig:ablation_radar} (bottom row) shows the same pattern: \textbf{w/o multi-hop} and \textbf{w/o tree} pull inward at L3-L5, while L1 stays near the full system. \textbf{w/o SelectLevel} mainly erodes the L5 spoke. Although \textbf{w/o enrichment} reduces the matched set most (26,250 pairs), it is listing, rather than fine-grained attribution, that suffers most without the facility index; conversely, \textbf{w/o multi-hop} leaves listing nearly unchanged yet sharply lowers matched L5 performance. Cross-section reading is thus dispensable for pair recall but essential for level assignment among listed pairs.

Overall, enrichment anchors inventory construction; multi-hop reading and tree indexing drive L3-L5 attribution in both modes; and SelectLevel mainly refines level boundaries, especially at L5.

\begin{table*}[th!]
\centering
\caption{Ablation results for FacilityAudit-Agent on the test set.
Left: \texttt{single\_audit} (n=39,358; L0-L5).
Right: \texttt{inventory\_audit} on the matched listed subset (n varies by variant; L1-L5).}
\label{tab:ablation_results}
\small
\setlength{\tabcolsep}{4.5pt}
\begin{tabular*}{\textwidth}{@{\hspace{1em}}l@{\extracolsep{\fill}}cccccccc@{\hspace{1em}}}
\hline\hline
& \multicolumn{4}{c}{\texttt{single\_audit}} & \multicolumn{4}{c}{\texttt{inventory\_audit}} \\
\cmidrule(l{-1pt}r{4pt}){2-5} \cmidrule(r{0pt}){6-9}
Variant & Acc & Macro-P & Macro-R & Macro-F1 & Acc & Macro-P & Macro-R & Macro-F1 \\
\midrule
w/o tree          & 81.0 & 64.1 & 71.4 & 66.6 & 84.4 & 66.2 & 75.4 & 69.1 \\
w/o enrichment    & 85.5 & 71.1 & 80.8 & 74.6 & 87.0 & 70.4 & 79.6 & 73.6 \\
w/o multi-hop     & 76.7 & 61.2 & 67.5 & 62.9 & 82.7 & 64.7 & 72.8 & 67.3 \\
w/o SelectLevel   & 84.9 & 70.2 & 79.6 & 73.6 & 86.6 & 69.6 & 78.8 & 72.7 \\
\textbf{FacilityAudit-Agent} & \textbf{89.1} & \textbf{75.9} & \textbf{85.7} & \textbf{79.8} & \textbf{89.8} & \textbf{75.3} & \textbf{83.9} & \textbf{78.6} \\
\hline
\end{tabular*}
\end{table*}

\subsection{Robustness across Models}\label{subsec:cross_model}

We next examine whether FacilityAudit-Agent depends on a particular LLM API. Keeping the indexing, enrichment, tool use, and SelectLevel steps fixed (Section~\ref{subsec:setup}), we rerun the full pipeline with five LLMs: GPT-5.5, GLM-5.2, DeepSeek-V4-Flash, Kimi K2.6, and Qwen3.6-27B. Unlike in Section~\ref{subsec:main_single}, we include no baseline methods here; each row in Table~\ref{tab:cross_model} refers to FacilityAudit-Agent.

Across all five models, FacilityAudit-Agent maintains strong performance, but the LLMs are not interchangeable in detail. Under \texttt{single\_audit} mode, macro-F1 spans 6.8 percentage points (75.6\%-82.4\%). GPT-5.5 achieves the best overall scores, followed by GLM-5.2. DeepSeek-V4-Flash reaches 89.1\% accuracy and 79.8\% macro-F1. Kimi K2.6 is nearly tied with DeepSeek-V4-Flash (within 0.2 percentage points in macro-F1), whereas Qwen3.6-27B ranks lowest yet still attains 87.1\% accuracy and 75.6\% macro-F1. Most of the Qwen3.6-27B gap concentrates at L3-L5; L1 remains nearly unchanged across LLMs.

The same ranking holds under \texttt{inventory\_audit} mode over the matched listed subset, with macro-F1 ranging from 74.4\% to 81.1\%. These results indicate that FacilityAudit-Agent's gains arise primarily from structured reading and level logic rather than from a single proprietary model, while stronger LLMs still provide a moderate additional boost at the L3-L5 boundary.

\begin{table*}[th!]
\centering
\caption{FacilityAudit-Agent with five LLM backends on the test set. Left: \texttt{single\_audit} (n=39,358; L0-L5). Right: \texttt{inventory\_audit} on the matched listed subset (n varies by LLM; L1-L5).}
\label{tab:cross_model}
\small
\setlength{\tabcolsep}{4.5pt}
\begin{tabular*}{\textwidth}{@{\hspace{1em}}l@{\extracolsep{\fill}}cccccccc@{\hspace{1em}}}
\hline\hline
& \multicolumn{4}{c}{\texttt{single\_audit}} & \multicolumn{4}{c}{\texttt{inventory\_audit}} \\
\cmidrule(l{-1pt}r{4pt}){2-5} \cmidrule(r{0pt}){6-9}
Model & Acc & Macro-P & Macro-R & Macro-F1 & Acc & Macro-P & Macro-R & Macro-F1 \\
\midrule
DeepSeek-V4-Flash  & 89.1 & 75.9 & 85.7 & 79.8 & 89.8 & 75.3 & 83.9 & 78.6 \\
GPT-5.5            & 90.8 & 78.6 & 87.8 & 82.4 & 91.3 & 77.8 & 86.0 & 81.1 \\
GLM-5.2            & 90.2 & 77.4 & 87.0 & 81.3 & 90.6 & 76.8 & 85.2 & 80.1 \\
Kimi K2.6          & 89.0 & 75.8 & 85.5 & 79.6 & 89.7 & 75.1 & 83.7 & 78.4 \\
Qwen3.6-27B        & 87.1 & 71.8 & 82.2 & 75.6 & 88.2 & 71.2 & 80.3 & 74.4 \\
\hline
\end{tabular*}
\end{table*}

\subsection{Error Analysis and Case Studies}\label{subsec:cases}

The confusion matrices in Figures~\ref{fig:single_matrix} and~\ref{fig:inventory_matrix} show where residual errors concentrate. For FacilityAudit-Agent, most off-diagonal mass lies between neighboring levels, especially L3\(\leftrightarrow\)L4 and L4\(\leftrightarrow\)L5, rather than long-range swaps such as L1\(\leftrightarrow\)L5. This pattern is consistent with the difficulty of deciding how deeply a facility's data enter the main argument once data use has already been established. L2 remains fragile for all methods because instrument-focused or engineering-focused papers are scarce in the test set, as also noted by \cite{wu2025amc}.

Under \texttt{inventory\_audit}, an additional failure mode appears before level scoring: facilities omitted from the predicted list never enter the matched matrix, so listing misses and level confusions should be read separately.

Two short cases illustrate the remaining challenges. In the first, a major facility is named repeatedly in the introduction and related-work sections, yet the analysis uses only archival data from another observatory. Keyword- and snippet-oriented baselines tend to treat the repeated mentions as evidence of scientific use, whereas FacilityAudit-Agent, after browsing the analysis and conclusion nodes, finds no data-use chain for the mentioned facility and assigns L1. In the second, sample selection cites facility \(f\) in the data section, but the dependence of the core results on \(f\) becomes explicit only in the discussion. Stopping at the first hit underestimates the contribution; linking the two sections supports an L4 or L5 decision. These examples mirror the ablation result that multi-hop reading matters most for L3-L5, while tree-guided navigation helps avoid mention-as-usage errors at L0-L1.

\subsection{Limitations}\label{subsec:limitations}

Several limitations follow from the present design and deployment setting.

First, FacilityAudit-Agent places comparatively high demands on the stability of the underlying LLM and of tool calling. The audit loop depends on well-formed JSON, grounded quotations, and successful tool returns within a capped number of rounds. Unstable generations (schema violations, invented quotations, repeated failed tool calls, or stalled tool responses) can break the pipeline, force retries, or leave the agent waiting until the round limit is reached. In practice, these failures make the audit slower and can stop the run before a usable level is returned. For deployment, the framework should therefore time out stalled calls, retry failed steps when safe, and send unfinished cases to human review.

Second, token consumption, and thus monetary cost, is substantial relative to single-pass baselines. The average token usage measured on the test papers is 273k tokens per paper for DeepSeek-V4-Flash, 142k for GPT-5.5, 195k for GLM-5.2, 206k for Kimi K2.6, and 455k for Qwen3.6-27B. The spread reflects differences in verbosity, tool-loop length, and how often the model re-fetches sections. Inventory mode is typically more expensive than single-facility mode because multiple facilities are audited in turn. Cost-aware deployments may therefore prefer a stronger but more concise LLM, cache tree indices across facilities on the same PDF, or restrict full multi-hop audits to pairs already flagged by a cheaper screen.

Third, the empirical scope remains bounded. The dataset covers ten facilities and a fixed set of journals and years; transfer to other missions, wavelengths, or publishers is untested. L2 is too rare for stable per-level conclusions. PDF quality (missing TOCs, scanned pages, publisher-specific layouts) can degrade tree construction before the agent ever runs.

Finally, FacilityAudit-Agent is intended as a scalable screening layer to assist in bibliography curation, rather than to replace the professional judgment of bibliographers. Predictions with low confidence, agent execution failures, and facility-name ambiguities should remain in the human review queue; the confidence indicators and conformal prediction flags described in Subsection~\ref{subsec:confidence} are designed for this workflow. Currently, the framework has been deployed at the National Astronomical Data Center of the National Astronomical Observatories, Chinese Academy of Sciences, and is applied to literature data processing for telescopes such as LAMOST, FAST, and EP. Its actual impact on curator time in real bibliography curation environments still awaits further assessment.

\section{Conclusion} \label{sec:conclusion}

Observatory bibliographies need more than a yes-or-no answer to whether a facility name appears in a paper. They need to know how that facility's data support the scientific argument. We cast this need as semantic contribution attribution on a six-level scale, L0-L5, and introduce FacilityAudit-Agent. The framework builds a page-bounded outline of each PDF, enriches sections with facility mentions, and uses a tool-calling language model to fetch text on demand. Evidence from data, analysis, and conclusion sections is combined through an explicit decision tree and a confidence score for human review.

On 79,407 human-annotated paper-facility pairs from 10 facilities and 10 journals (39,358 in the test set), FacilityAudit-Agent reaches 89.1\% accuracy and 79.8\% macro-F1 in single-facility mode, outperforming full-context LLM prompting, chunk-based retrieval, and a TRACS-style classifier, with the clearest gains at L2 and L5. In inventory mode, it also leads on facility listing and on level assignment among listed pairs. Ablation shows that multi-hop reading and tree indexing matter most for L3-L5 attribution, while enrichment mainly helps inventory construction and SelectLevel mainly sharpens L5 boundaries. Across five LLM backends, performance stays strong, with moderate gaps concentrated at high contribution levels.

Remaining errors are mostly swaps between neighboring levels. The agent also needs stable tool calling, and token cost is high relative to single-pass baselines, so deployment should use timeouts, retries, and a review queue. FacilityAudit-Agent is best viewed as a screening layer for bibliography curation, not a replacement for expert judgment. It is already in use at the National Astronomical Data Center (NADC) for literature related to facilities such as LAMOST, FAST, and EP. Future work includes broader facility coverage, lower-cost audits, and measuring how much curator time the review flags save.

\section*{Acknowledgements}

This work is supported by the National Key R\&D Program of China (2022YFF0711500), National Natural Science Foundation of China (NSFC) 12273077, 12403102, 12373110, 12103070, and 1240030222, Strategic Priority Research Program of the Chinese Academy of Sciences (XDB0550101). Data resources are supported by China National Astronomical Data Center (NADC), CAS Astronomical Data Center and Chinese Virtual Observatory (China-VO). This work is supported by Astronomical Big Data Joint Research Center, co-founded by National Astronomical Observatories, Chinese Academy of Sciences and Alibaba Cloud.

\appendix
\section{Abbreviations of Facility Names} \label{app:abbreviation}
Table~\ref{tab:facility_aliases} lists the ten facilities used to build the dataset in Section~\ref{sec:data}. For each facility we show the abbreviation used in this paper and the full name.

\begin{table*}[th!]
\centering
\caption{Abbreviations of the facilities used in this paper and their full names.}
\label{tab:facility_aliases}
\begin{tabular*}{\textwidth}{@{\hspace{1em}}l@{\extracolsep{\fill}}l@{\hspace{2em}}l@{\extracolsep{\fill}}l@{\hspace{1em}}}
\hline\hline
Abbreviation & Full name & Abbreviation & Full name \\
\hline
LAMOST & Large Sky Area Multi-Object Fiber Spectroscopic Telescope & EP & Einstein Probe \\
FAST & Five-hundred-meter Aperture Spherical radio Telescope & VLT & Very Large Telescope \\
Keck & W.\ M.\ Keck Observatory & SDSS & Sloan Digital Sky Survey \\
GTC & Gran Telescopio Canarias & HST & Hubble Space Telescope \\
Subaru & Subaru Telescope & Gemini & Gemini Observatory \\
\hline
\end{tabular*}
\end{table*}

\section{Cohen's $\kappa$} \label{app:cohen}

To check how consistently human annotators apply the L0-L5 scale, we drew a subset of 2,000 paper-facility pairs and asked two annotators to label each pair independently. Both annotators used the same level definitions (Table~\ref{tab:levels}) and read the full paper text. They did not see each other's labels. Agreement was measured with Cohen's $\kappa$ over the six levels:
\begin{equation}
\kappa=\frac{p_o-p_e}{1-p_e}.
\end{equation}
Here \(p_o\) is the observed agreement, that is, the fraction of pairs on which the two annotators assign the same level:
\begin{equation}
p_o=\sum_{i=0}^{5} p_{ii},
\end{equation}
where \(p_{ii}\) is the fraction of pairs labeled \(\mathrm{L}i\) by both annotators. The quantity \(p_e\) is the agreement expected by chance if the two annotators chose labels independently with their own marginal rates:
\begin{equation}
p_e=\sum_{i=0}^{5} p_{i\cdot}\,p_{\cdot i},
\end{equation}
where \(p_{i\cdot}\) is the fraction of pairs labeled \(\mathrm{L}i\) by the first annotator and \(p_{\cdot i}\) is the fraction labeled \(\mathrm{L}i\) by the second annotator. Thus \(\kappa=1\) means perfect agreement, \(\kappa=0\) means agreement no better than chance, and negative values mean agreement worse than chance. We obtain \(\kappa=0.87\), which indicates strong agreement for a six-way label. When the two labels differed, a third annotator reviewed the pair and chose the final label used in the dataset.

\bibliographystyle{elsarticle-harv}
\bibliography{sample701}

\end{document}